%% file: main.tex
\documentclass[%
reprint, 
superscriptaddress,
 amsmath,amssymb,
 aps, physrev,
pra,
]{revtex4-2}

\usepackage{graphicx}
\usepackage{dcolumn}
\usepackage{float}

\usepackage{refcount}

\usepackage{bm}

\begin{document}

\preprint{APS/123-QED}

\title{Strong Coupling Between RF Photons and Plasmons of Electrons on Liquid Helium}
\author{Asher Jennings}
\email{asher.jennings@riken.jp}
\affiliation{Center for Quantum Computing, RIKEN, Wako, Saitama 351-0198, Japan}

\author{Ivan Grytsenko}
\affiliation{Center for Quantum Computing, RIKEN, Wako, Saitama 351-0198, Japan}

\author{Thomas Giovansili}
\affiliation{Center for Quantum Computing, RIKEN, Wako, Saitama 351-0198, Japan}
\affiliation{\'Ecole Polytechnique, Route de Saclay, Palaiseau Cedex 91128, France}

\author{Itay Josef Barabash}
\affiliation{Center for Quantum Computing, RIKEN, Wako, Saitama 351-0198, Japan}
\affiliation{School of Computer Science, McGill University, Montreal, Quebec H3A 2K6, Canada}

\author{Oleksiy Rybalko}
\affiliation{Center for Quantum Computing, RIKEN, Wako, Saitama 351-0198, Japan}
\affiliation{Physics of Quantum Fluids and Crystals, B. Verkin Institute for Low Temperature Physics and Engineering of the National Academy of Sciences of Ukraine, Kharkiv 61103, Ukraine}

\author{Yiran Tian}
\affiliation{Center for Quantum Computing, RIKEN, Wako, Saitama 351-0198, Japan}
\affiliation{Institute of Physics, Kazan Federal University, Kazan 420008, Republic of Tatarstan, Russian Federation}

\author{Jung Wang}
\affiliation{Center for Quantum Computing, RIKEN, Wako, Saitama 351-0198, Japan}

\author{Hiroki Ikegami}
\affiliation{Center for Quantum Computing, RIKEN, Wako, Saitama 351-0198, Japan}
\affiliation{Beijing National Laboratory for Condensed Matter Physics, Institute of Physics, Chinese Academy of Sciences, Beijing 100190, China}

\author{Erika Kawakami}
\email{e2006k@gmail.com}
\affiliation{Center for Quantum Computing, RIKEN, Wako, Saitama 351-0198, Japan}
\affiliation{Pioneering Research Institute, RIKEN, Wako, Saitama 351-0198, Japan}

\date{\today}

\begin{abstract}
Plasmons, arising from the collective motion of electrons, can interact
strongly with electromagnetic fields or photons; this capability has been exploited
across a broad range of applications, from chemical reactivity to biosensing. Recently, there has been growing interest in plasmons for applications in
quantum information processing. Electrons floating on liquid helium provide an
exceptionally clean, disorder-free system and have emerged as a promising
platform for this purpose. In this work, we establish this system as a tunable
plasmon–photon hybrid platform. We demonstrate strong coupling between floating-electron plasmons and radio-frequency (RF) photons confined in an LC resonator.
Time-resolved measurements reveal coherent oscillatory energy exchange between the plasmonic and photonic modes, providing direct evidence of their coherent coupling. These results represent a step towards cavity quantum electrodynamics with a floating-electron plasmon coupled to a resonator. Furthermore, the LC resonator serves as a sensitive probe of electron-on-helium physics, enabling the observation of the Wigner crystal transition and a quantitative study of the temperature-dependent plasmon decay arising from ripplon-induced scattering.
\end{abstract}

\maketitle


\section{Introduction}\label{sec:intro}
Longitudinal collective oscillations of electrons, referred to as plasmons, are fundamental excitations originating from long-range Coulomb interactions in diverse electronic systems, ranging from metallic particles to two-dimensional (2D) electron systems. When these oscillations are spatially confined, they are referred to as localized plasmons. The resulting strong electromagnetic field confinement enhances light--matter interactions, enabling a variety of plasmonic applications. In metallic nanoparticles, electrons oscillate coherently in response to an incident optical field, giving rise to localized surface plasmons~\cite{Tame2013-tp,Liu2023-sw}. These effects have been exploited for photocatalysis~\cite{Serpone2012-bs}, plasmon-assisted photovoltaics~\cite{Clavero2014-yd}, photodetection~\cite{Knight2011-ou}, and biosensing~\cite{Shrivastav2021-bq}. 

In 2D electron systems in semiconductor materials, plasmons have recently attracted growing interest as carriers of quantum information. The coherent generation and manipulation of plasmon wave packets offers a promising route toward transmitting quantum information over finite distances, as proposed and explored in recent theoretical and experimental studies~\cite{Braunstein2005-yc,Takada2025-sj,Yoshioka2024-wj,Gonzalez-Tudela2011-ty}. Such propagating modes could, in principle, be coherently captured and stored in a localized cavity mode, providing a means to interface mobile and stationary
quantum excitations. To route quantum information beyond the 2D electron system the ability to convert plasmonic states into photons is required.
To achieve this, it is of prime importance to realize the strong-coupling regime—where the coherent coupling rate exceeds the dissipation rates. 

However, in both metals and 2D semiconductor systems, plasmonic dissipation losses typically make access to this regime challenging, although a notable exception has been demonstrated using periodic arrays of metal nanoparticles~\cite{Mueller2020-ne}, where the enhanced plasmon--photon coupling exceeds the plasmon losses.

In this work, we overcome this limitation by employing floating electrons in vacuum, an intrinsically low-dissipation electronic system that enables strong coupling between localized plasmons and photons. Electrons float in vacuum approximately \(10~\mathrm{nm}\) above the liquid helium and form a 2D electron system~\cite{Cole1969-my,shikin1970motion,Monarkha2004-un,
Andrei1997Two-DimensionalSubstrates,Crandall1972-hj}. Owing to the absence of
disorder and impurities in this pristine environment, electron scattering—and
thus dissipation—is strongly suppressed. This exceptionally clean environment has motivated recent proposals and experiments
to use single electrons on helium as qubits%
~\cite{Platzman1999,Koolstra2019-mq,Zhou2022-nk,Zhou2023-iw,Jennings2024-sb}. Most recently, strong coupling between the orbital state of a single electron on liquid helium and a microwave photon confined in a superconducting resonator has been demonstrated~\cite{Koolstra2025-qt}. In addition to single-electron dynamics, plasmons were first detected several decades ago using RF absorption techniques in the 50–200 MHz range~\cite{Grimes1976-se,Grimes1976-kr}. Since then, plasmons have not only been studied for their intrinsic properties~\cite{Andrei1994,Glattli1985-fp}, but have also been used as a probe to discover the Wigner crystal~\cite{Grimes1976-kr} and to investigate the surface properties of superfluid $^3$He~\cite{Kirichek1998-xy}. Recently, microwave plasmon spectroscopy in the few-gigahertz
range has been demonstrated through electron transport measurements%
~\cite{Mikolas2025-fb}.  Here, we demonstrate strong coupling between a plasmon mode of electrons on liquid helium and RF photons confined in an LC resonator (Fig.~\ref{fig1}(a)). This plasmon--photon coupling 
platform offers a high degree of tunability, as the electron density, coupling
strength, and temperature can be precisely controlled. Combined with the
low dissipation enabled by the pristine electron environment, this allows
systematic studies of plasmon--photon interactions, including
real-time coherent energy exchange in a regime difficult to access in
conventional semiconductor 2D materials. Beyond this, we demonstrate that the LC resonator serves as a sensitive probe of electron-on-helium physics by observing the Wigner crystal transition and determining the temperature dependence of the plasmon decay rate.

\section{Device}\label{sec:device}

\begin{figure*}
    \centering
    \includegraphics[width=0.7\linewidth]{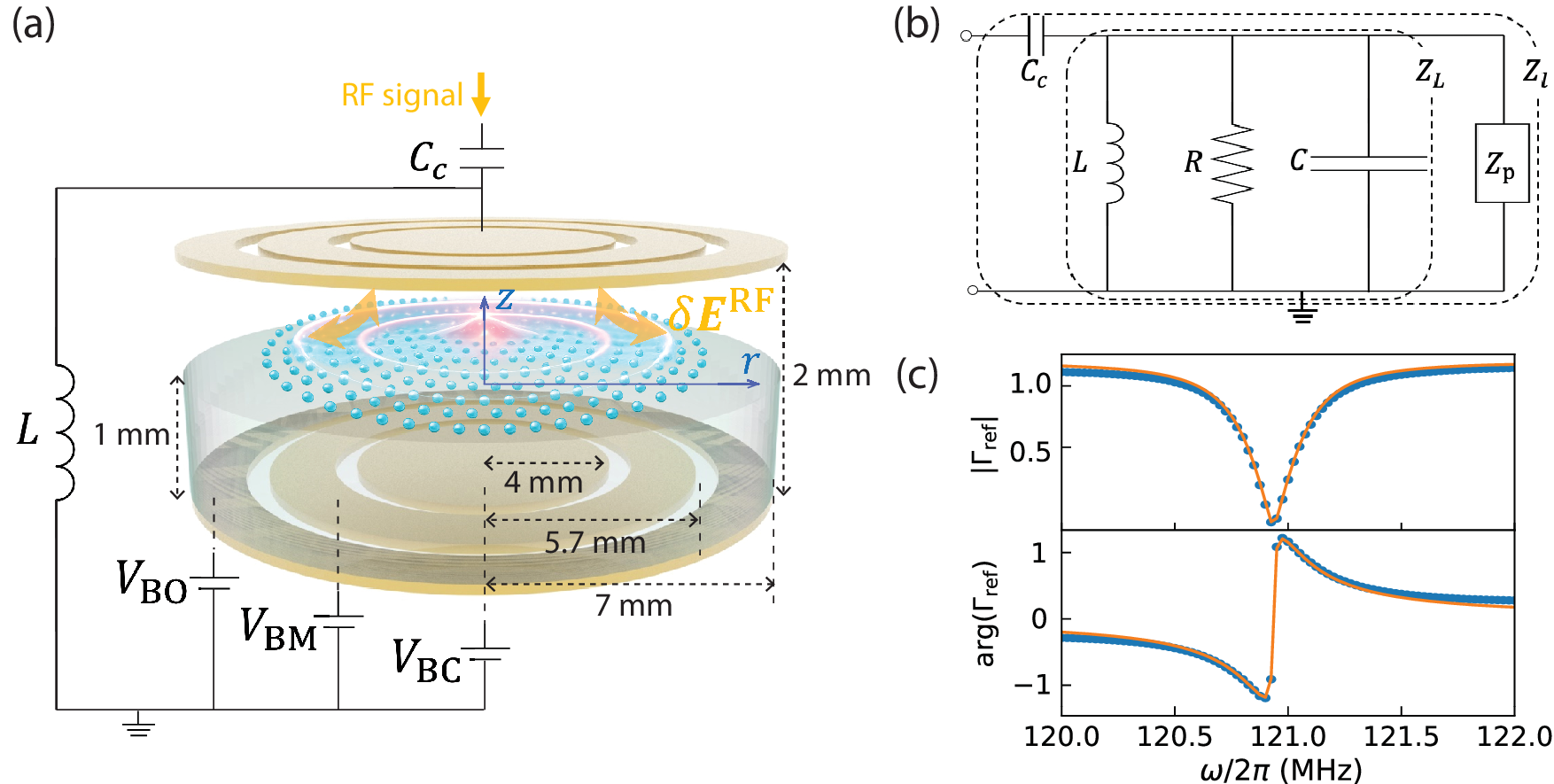}
    \caption{(a) Experimental setup with Corbino-geometry electrodes forming part of an LC resonator. The device consists of radially arranged top and bottom electrodes, each with an
area of approximately \(0.5~\mathrm{cm}^2\), which form the main capacitance
\(C\) of the LC resonator. The top center electrode is connected to an inductor
\(L\), which provides the resonant inductance, and to a coupling capacitor
\(C_\mathrm{C}\) that interfaces the resonator with the external circuit. The bottom electrodes are immersed in a 1-mm-thick layer of liquid helium (light
blue cylinder), above whose surface electrons (light blue circles) float approximately \(10~\mathrm{nm}\)
in vacuum. An RF signal applied to the top center electrode generates an oscillating radial electric field $\delta E_r^\mathrm{RF}$ confined within the LC resonator and drives collective electron oscillations (plasmon mode, schematically illustrated by the pink wave) in electrons on helium. (b) Schematic representation of the experimental setup in the form of an equivalent electrical circuit. The plasmon excitation is modeled by an effective impedance \( Z_\mathrm{p} \). Further details regarding the circuit elements are provided in the main text. (c) Measured magnitude and phase of the reflection coefficient of the LC resonant circuit, $|\Gamma_\mathrm{ref}|$ and $\arg(\Gamma_\mathrm{ref})$, in the absence of electrons (blue dots). The orange lines show a fit to Eq.~\eqref{eq:reflection_coefficient_input_output} with $g=0$. The data shown here are corrected for the phase offset and cable delay and are normalized, as is the case for all reflection data presented in this manuscript.}
    \label{fig1}
\end{figure*}

We realize plasmons of 2D electrons floating on liquid
helium confined within a circular area and couple them to
the electromagnetic mode of an LC resonator (Fig.~\ref{fig1}(a)). To laterally confine the electrons 
region, we employ a so-called Corbino geometry~\cite{Iye1980-zn} consisting of two parallel plates. Each plate comprises three concentric electrodes--- center, middle, and outer -- and we apply DC voltages \( V_\mathrm{BC} \),
\( V_\mathrm{BM} \), and \( V_\mathrm{BO} \), to the bottom electrodes respectively, while the electrodes of the top plate are left DC grounded. By
adjusting these voltages, the radial extent of the region in which electrons
can be stably confined is controlled. A nanofabricated inductor is
connected to the top center electrode to form the LC resonator~%
\cite{Jennings2025-ml}. The separation between the top and bottom plates is \( D = 2~\mathrm{mm} \). Liquid helium with a thickness of \( 1~\mathrm{mm} \) fills half of this
gap. Electrons are deposited onto the liquid helium surface via thermionic
emission~\cite{Williams1971-pg,Jennings2024-sb} and subsequently float
approximately \( 10~\mathrm{nm} \) above the liquid helium surface.

We measure the reflection of an RF signal applied to the top center electrode.
Fig.~\ref{fig1}(b) shows the equivalent electrical circuit of the measurement
setup. The reflection coefficient is given by
\begin{equation}
    \Gamma_\mathrm{ref}
    = \frac{Z_l - Z_0}{Z_l + Z_0},
    \label{eq:Ref_coeff}
\end{equation}
where \( Z_0 \) is the characteristic impedance of the transmission line and
\( Z_l \) is the load impedance which represents the coupled system of the LC resonator and plasmons~\cite{note:minus_sign},
\begin{equation}
    Z_l
    = \frac{1}{-i\omega C_\mathrm{c}}
    + \left( \frac{1}{Z_L} + \frac{1}{Z_p} \right)^{-1}.
    \label{eq:Z_l_delta_omega}
\end{equation}
Here \( C_\mathrm{c} = 0.3~\mathrm{pF} \) is the coupling capacitor and $i=\sqrt{-1}$. The impedance of the parallel LCR circuit is given by $   \frac{1}{Z_L}
    = \frac{1}{R}
    - i\omega C
    - \frac{1}{i\omega L}$, 
while the plasmon impedance \( Z_p \) is defined later. The circuit parameters are
\( L = 708~\mathrm{nH} \),
\( C = 2.131~\mathrm{pF} \), and
\( R = 321~\mathrm{k}\Omega \). The resonance frequency is \(
    \omega_0 / 2 \pi
    = 1/2 \pi\sqrt{L C_\mathrm{t}}
    = 120.946~\mathrm{MHz},
\) where \( C_\mathrm{t} = C + C_\mathrm{c} \).
The device used here is identical to that reported in
Ref.~\onlinecite{Jennings2025-ml}, where further circuit details are provided. Within the framework of input--output theory, Eq.~\eqref{eq:Ref_coeff} can be
rewritten as (Supplementary Information~\ref{sec:ref_coeff})
\begin{equation}
    \Gamma_\mathrm{ref}
    = 1 - \frac{-i\kappa_\mathrm{ext}}{
    (\omega_0 - \omega)
    - i\frac{\kappa_\mathrm{ext} + \kappa_\mathrm{int}}{2}
    + \dfrac{g^2}{(\omega - \omega_\mathrm{p}) + i\gamma_\mathrm{p}/2}},
    \label{eq:reflection_coefficient_input_output}
\end{equation}
where \( \omega_\mathrm{p} \) is the angular frequency of the plasmon mode,
\( g \) is the coupling strength between the resonator and the plasmon mode,
\( \kappa_\mathrm{ext} /2\pi=0.19\)~MHz and \( \kappa_\mathrm{int}/2\pi=0.20 \)~MHz are the external and
internal decay rates of the resonator, respectively, and
\( \gamma_\mathrm{p} \) is the decay rate of the plasmon mode. In the absence of electrons (\( g = 0 \)),
Eq.~\eqref{eq:reflection_coefficient_input_output} reduces to the standard
expression for a single-port resonator.
The corresponding measured reflection coefficient is shown in
Fig.~\ref{fig1}(c).

\section{Frequency-domain measurement}
The radial component of the RF field induced by the incident signal on the top
center electrode drives charge-density oscillations in the electron layer
(Fig.~\ref{fig1}(a)). This leads to coupling between the confined plasmon mode
and the resonator field when the plasmon mode frequency $\omega_{\mathrm{p}}$
approaches the resonator frequency $\omega_0$. The angular frequency of a 2D plasmon mode for a uniform electron density confined between the top and bottom electrodes, is given by~\cite{Glattli1985-fp,Ott-Rowland1982-bz}
\begin{equation}
    \omega_\mathrm{p} = \sqrt{ \frac{e^2 n_0 k_{\nu,\mu}}{2 m_e \varepsilon_0} \tanh\left( \frac{k_{\nu,\mu} D}{2} \right) },
    \label{eq:omega_p}
\end{equation}
where \( e \) is the elementary charge, \( n_0 \) is the 2D electron density, \( m_e \) denotes the free electron mass, \( \varepsilon_0 \) is the vacuum permittivity.  Here the relative permittivity of liquid helium is approximated as $\varepsilon_\mathrm{r}=1$. The quantity \( k_{\nu,\mu} \) represents the wave number of the plasmon mode characterized by the azimuthal mode index \( \nu \) and the radial mode index \( \mu \) in the cylindrical geometry. Due to our device geometry, only radial modes are efficiently driven; therefore,
we consider only the fundamental azimuthal mode (\(\nu = 0\)). Note that the wave number $k_{0,\mu}$ is determined by the boundary condition at the edge of the circular electron sheet and therefore depends on its radius.

To prepare electrons on the helium surface, we first set a positive voltage to all the bottom electrodes $ V_{\mathrm{BC}} = V_{\mathrm{BM}} = V_{\mathrm{BO}} > 0$.
Electrons are then deposited onto the helium surface until the saturation density
is reached, corresponding to the maximum electron density allowed for the given
$V_{\mathrm{BC}}$. We confine the electrons primarily to the center electrode by first sweeping the outer electrode to $V_{\mathrm{BO}} = -32~\mathrm{V}$, and then the middle electrode $V_{\mathrm{BM}} = -32~\mathrm{V}$.

In frequency-domain measurements, the plasmon frequency $\omega_{\mathrm{p}}$
is tuned by simultaneously adjusting the electron density $n_0$ and the wave vector $k_{0,\mu}$ (see Eq.~\ref{eq:omega_p}).  \( V_\mathrm{BM} \) was subsequently swept
between \(-32~\mathrm{V}\) and \(5~\mathrm{V}\) while keeping \( V_\mathrm{BO} \) and
\( V_\mathrm{BC} \) constant.

\begin{figure*}
    \centering
    \includegraphics[width=\textwidth]{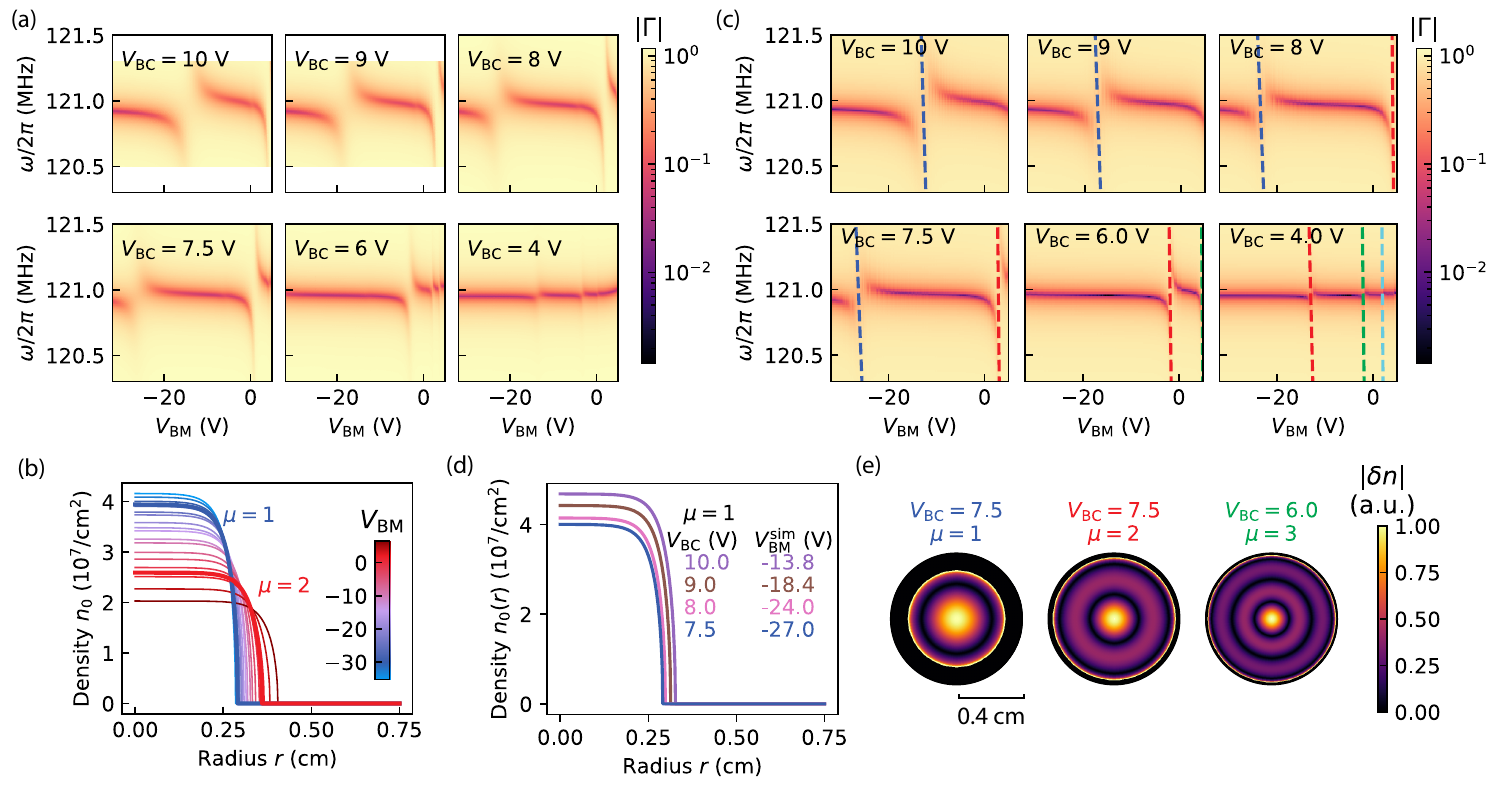}
    \caption{Magnitude of the reflection coefficient \( |\Gamma| \) as a function of $V_\mathrm{BM}$ and RF signal frequency $\omega/2\pi$ from both experiment in (a) and simulation in (c) for different $V_\mathrm{BC}=10, 9, 8 , 7.5, 6, 4$ V. $V_\mathrm{BO}$ is fixed to $-32$~V. (a) Measurement temperature $T = 180~\mathrm{mK}$; The RF excitation power is set to \(-50~\mathrm{dBm}\) at the output of the signal generator. (b) Simulated electron density profiles $n_0(r)$ for different values of
$V_\mathrm{BM}$, with $V_\mathrm{BC}$ fixed at $7.5~\mathrm{V}$ and $V_\mathrm{BO}$ fixed at $-32~\mathrm{V}$. For all values of $V_\mathrm{BM}$, the total number of electrons is kept constant. The thick blue and red lines correspond to the density profiles for which the
fundamental and second plasmon modes, respectively, are tuned into resonance
with the LC resonator frequency according to the simulation shown in (c).
 (c) Simulated magnitude of the reflection coefficient. Overlaid dashed lines indicate the calculated plasmon-mode frequencies: the fundamental (\(\mu = 1\), blue), second (\(\mu = 2\), red), third (\(\mu = 3\), green), and fourth (\(\mu = 4\), light blue) modes. 
 (d) Simulated electron density profiles $n_0(r)$ for $V_\mathrm{BC}=10, 9, 8,$ and $7.5~\mathrm{V}$, evaluated at $V_\mathrm{BM}^{\mathrm{sim}}$ where the fundamental plasmon mode ($\mu=1$) is resonant with the LC resonator according to the simulation shown in (c). (e) Simulated magnitude of the density modulation, \( |\delta n|\), at resonance for the fundamental (\( \mu = 1 \)) 
and second (\( \mu = 2 \)) modes at \( V_\mathrm{BC} = 7.5~\mathrm{V} \), and for the third mode (\( \mu = 3 \)) at 
\( V_\mathrm{BC} = 6~\mathrm{V} \). The plasmon decay rates used in the simulations can be found in Supplementary Information~\ref{sec:ref_coeff}, Fig.~\ref{fig:Fig_gamma_field}.
}
    \label{fig2}
\end{figure*}

The measured reflection coefficient is shown in Fig.~\ref{fig2}(a) for several
different values of $V_\mathrm{BC}$ (i.e., several different saturation densities). When $\omega_\mathrm{p}$ is tuned into resonance with the LC
resonator ($\omega_p=\omega_0$), an avoided level crossing appears. The electron density profiles
for several values of $V_\mathrm{BM}$ at fixed $V_\mathrm{BC} = 7.5~\mathrm{V}$
and $V_\mathrm{BO} = -32~\mathrm{V}$ are shown in Fig.~\ref{fig2}(b). These
profiles are obtained by solving the Poisson equation for the actual device
geometry, using the applied voltages $V_\mathrm{BO}$, $V_\mathrm{BM}$, and
$V_\mathrm{BC}$ as input parameters~\cite{Wilen1988-ui} (see
Refs.~\onlinecite{Jennings2025-ml,density_profile_repo} for details of the numerical
procedure). During this $V_\mathrm{BM}$ sweep,  the electron pool radius
and the electron density $n_0$ are varied simultaneously while keeping the
total number of electrons fixed, thereby tuning $\omega_\mathrm{p}$. As seen in Fig.~\ref{fig2}(a) for \(V_\mathrm{BC}=7.5~\mathrm{V}\), avoided crossings occur at
\(V_\mathrm{BM}^\mathrm{exp}=-26.4\) ~V and 0.8~V.
These avoided crossings occur when the plasmon frequencies \(\omega_\mathrm{p}\) of the
fundamental and second modes are tuned to the resonator frequency \(\omega_0\),
respectively, as determined from the \(\omega_\mathrm{p}\) calculations using the
procedure described below.

For a quick estimate of \(\omega_\mathrm{p}\), we adopt the following approximation.
We model the electron sheet as a disk of uniform electron density \(n_0\) with an
effective radius \(R^\ast\).
The radius \(R^\ast\) is defined as the radius at which the simulated density
drops to zero, and \(n_0\) is chosen such that the total number of electrons in
the disk is equal to that obtained from the simulated density profile.
The wave number \(k_{0,\mu}\) is determined by the boundary condition that the
radial current vanishes at \(R^\ast\), which is given by
\(
J_0'(k_{0,\mu} R^\ast)=0,
\)
where \(J_0'\) denotes the derivative of the zeroth-order Bessel function%
~\cite{Glattli1985-fp} (see Supplementary Information~\ref{sec:plasmon_freq}). Using this approximation, we obtain a fundamental (\(\mu=1\)) plasmon frequency
of \(\omega_\mathrm{p}/2\pi=\) 125.0~MHz for \(V_\mathrm{BC}=7.5~\mathrm{V}\) and
\(V_\mathrm{BM}^\mathrm{exp}= -26.4\)~V.
We therefore attribute the observed avoided crossing at this bias point to the
fundamental plasmon mode being tuned into resonance with the LC resonator.
The deviation from \(\omega_0\) can be attributed to the
uniform-density approximation, which neglects the gradual decrease of the
electron density near the edge of the electron sheet
(Supplementary Information~\ref{sec:plasmon_freq}). The second mode ($\mu = 2$) plasmon frequency at the avoided crossing observed at
$V_\mathrm{BM}^\mathrm{exp} = 0.8~\mathrm{V}$ is estimated to be
$\omega_\mathrm{p}/2\pi \simeq 130.5~\mathrm{MHz}$, which is noticeably higher than the bare resonator frequency
$\omega_0/2\pi$. A definitive mode assignment cannot be made from this quick estimate alone. As discussed below, a more accurate evaluation shows that this feature indeed originates from the second plasmon mode.

A more accurate estimate of $\omega_\mathrm{p}$ is obtained by treating
them as the eigenvalues of a self-consistent charge-density equation derived
from the continuity equation and Coulomb interactions (see
Ref.~\onlinecite{Wilen1988-ui} and Sec.~\ref{sec:Green_Sim} in Methods), using
only the simulated density profiles as input. The resulting plasmon frequencies for the first through fourth radial modes are shown in Fig.~\ref{fig2}(c) as dashed lines (blue, orange, green, and light blue). Fig.~\ref{fig2}(d) shows the electron density profiles when the fundamental
plasmon mode frequency is tuned to the resonator frequency for four different
values of \(V_\mathrm{BC}\). This approach also yields the plasmon impedance \( Z_\mathrm{p} \), defined through the charge variation \( \delta Q \) induced on the top center electrode by the plasmonic density oscillation, quantifying how the plasmon mode couples to the resonator  (Sec.~\ref{sec:Green_Sim} in Methods). The simulation results of the reflection coefficient, calculated using Eq.~\ref{eq:Z_l_delta_omega} with the corresponding plasmon impedance \( Z_\mathrm{p} \), are shown in Fig.~\ref{fig2}(c). For the reflection-coefficient simulations, the plasmon relaxation rate
$\gamma_\mathrm{p}$ is included as an additional input parameter and its value is
extracted from the experimental data (Supplementary Information~\ref{sec:ref_coeff}).
The simulation results in Fig.~\ref{fig2}(c) are in good agreement with the experimental data in Fig.~\ref{fig2}(a), both in the resonance positions and in the coupling strength between the LC resonator field and the plasmon mode. 
A comparison of these two figures shows, for example, that the observed
avoided crossing at \(V_\mathrm{BC} = 7.5~\mathrm{V}\) and
\(V_\mathrm{BM}^\mathrm{exp} = 0.8~\mathrm{V}\) in Fig.~\ref{fig2}(a) can be attributed to the second
(\(\mu = 2\)) plasmon mode being tuned into resonance with the LC resonator.

The remaining deviations between experiment and simulation, particularly for positive $V_\mathrm{BM}$, may arise from
geometrical elements not included in the simulation, e.g., the gap between the electrodes. The magnitude of the density modulation for the first, second, and third radial plasmon modes, $|\delta n(r)|$, evaluated at resonance, is shown in Fig.~\ref{fig2}(e). The number and positions of the nodes directly illustrate how standing waves are formed in each radial mode.

\section{Strong coupling and Time-Domain Dynamics}\label{sec:time-domain}
To enhance the coupling strength $g$ and enter the strong-coupling regime, we
populate electrons above both the bottom-center and bottom-middle electrodes by
setting $V_\mathrm{BC}=V_\mathrm{BM}=17~\mathrm{V}$ while applying a negative
bias $V_\mathrm{BO}$. This configuration extends the electron cloud toward the
outer region of the top-center electrode ($r \simeq 4~\mathrm{mm}$), where the
radial RF electric field $\delta E_r^{\mathrm{RF}}$ is strongest. As a result,
the spatial overlap between the plasmon current and the resonator field is
enhanced, thereby increasing the coupling strength $g$
(Supplementary Information~\ref{sec:coupling_g}).

\begin{figure}[h!]
    \centering
\includegraphics[width=1\linewidth]{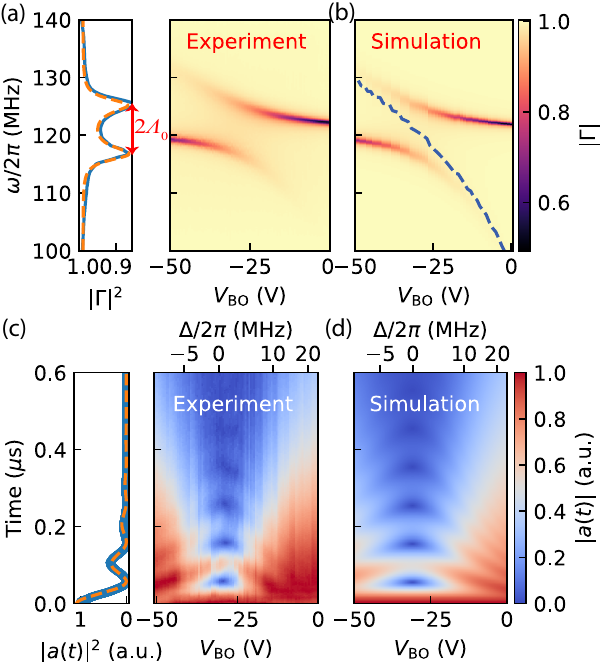}
    \caption{(a,c) Measurement temperature $T = 180~\mathrm{mK}$; The RF excitation power is set to \(-20~\mathrm{dBm}\) at the output of the signal generator.
 (a) Right: Normalized measured reflection coefficient as a function of the bottom
outer electrode voltage $V_\mathrm{BO}$ and the RF frequency applied to the LC resonator. The other electrode voltages are fixed at
$V_\mathrm{BC} = V_\mathrm{BM} = 17~\mathrm{V}$. An avoided crossing is observed when the plasmon frequency is tuned into resonance with the LC circuit, indicating coherent coupling between the two modes. Left: Reflection spectrum at the resonance point $V_\mathrm{BO} = -29~\mathrm{V}$ (blue solid line), together with a fit to
Eq.~\ref{eq:reflection_coefficient_input_output} (orange dashed line). The fit yields a coupling strength
$g/2\pi = 4.55 \pm 0.02~\mathrm{MHz}$  and a plasmon decay rate
$\gamma_\mathrm{p}/2\pi = 5.10 \pm 0.07~\mathrm{MHz}$ . From these parameters, we obtain a mode splitting
$2\Lambda_0\approx 9.0~\mathrm{MHz}$ (Eq.~\ref{eq:Lambda}).
(b) Numerically simulated reflection coefficient corresponding to panel (a).
The plasmon decay rate $\gamma_{\mathrm p}/2\pi = 5.1~\mathrm{MHz}$ is used as an
input parameter for the simulation. At the resonance voltage
$V_\mathrm{BO} = -30.7~\mathrm{V}$, fitting the simulated spectrum yields
$g/2\pi = 4.55~\mathrm{MHz}$, consistent with the experiment. The overlaid dashed blue lines indicate the calculated fundamental plasmon
mode frequencies. (c) Measured $\sqrt{I^2 + Q^2}$ of the reflected signal following a
$20~\mathrm{ns}$ RF pulse.
Inset: Time-resolved $I^2 + Q^2$ response measured at the resonance point
$V_\mathrm{BO} = -29~\mathrm{V}$ (blue dots).
The orange dashed line shows a fit to Eq.~\ref{eq:a_square}, up to an overall
proportionality constant, yielding
$g/2\pi = 4.906 \pm 0.009~\mathrm{MHz}$ and
$\gamma_{\mathrm p}/2\pi = 3.30 \pm 0.02~\mathrm{MHz}$. (d) Analytical calculation corresponding to panel (c), performed with the
resonance point set to $V_\mathrm{BO} = -29~\mathrm{V}$.
The color scale represents the amplitude of the RF field stored in the LC
resonator, $|a(t)|$, calculated using Eq.~\ref{eq:a_general} with
$g/2\pi = 4.9~\mathrm{MHz}$,
$\gamma_{\mathrm p}/2\pi = 3.3~\mathrm{MHz}$, and
$\kappa/2\pi = 0.39~\mathrm{MHz}$. (c,d) The detuning on the upper horizontal axis,
$\Delta = \omega_0 - \omega_\mathrm{p} $, is obtained from the numerically calculated relation between $V_{\mathrm{BO}}$ and $\omega_\mathrm{p}$.}
\label{fig:strong_coupling_time_domain}
\end{figure}

In this experiment, \(V_\mathrm{BO}\) is swept, rather than \(V_\mathrm{BM}\), to tune \(\omega_\mathrm{p}\). Fig.~\ref{fig:strong_coupling_time_domain}(a) shows the reflected signal as a
function of \(V_\mathrm{BO}\).
At \(V_\mathrm{BO} = -29~\mathrm{V}\), where the plasmon mode is tuned into resonance
with the LC resonator, a clear splitting of the resonance peak is observed,
indicating hybridization between the plasmon and resonator modes.
 Fitting the spectrum with
Eq.~\eqref{eq:reflection_coefficient_input_output}
 yields $g/2\pi = 4.55 \pm 0.02~\mathrm{MHz}$ and $\gamma_\mathrm{p}/2\pi = 5.10 \pm 0.07~\mathrm{MHz}$ (Supplementary Information~\ref{sec:ref_coeff}). The mode splitting at resonance is given by $2\Lambda_0$ (Supplementary Information~\ref{sec:ref_coeff}), where 
\begin{equation}
\Lambda_0 = \sqrt{\,g^2 - \left(\frac{\Delta_\kappa}{4}\right)^2\,},
\label{eq:Lambda}
\end{equation}
with $\Delta_\kappa = \kappa - \gamma_\mathrm{p}$. $\kappa/2\pi = (\kappa_\mathrm{int} + \kappa_\mathrm{ext})/2\pi
= 0.39~\mathrm{MHz}$ is the total resonator decay rate. The observation of a real-valued $\Lambda_0$ and the associated peak splitting is a hallmark of the strong-coupling regime. The corresponding simulation is shown in Fig.~\ref{fig:strong_coupling_time_domain}(b) and exhibits good agreement with the measured spectrum in terms of both the resonance position and the coupling strength.

Furthermore, to directly probe the dynamical coupling between the LC resonator and the plasmon mode, we performed time-domain microwave reflectometry measurements. A 20~ns RF pulse was injected into the resonator, and the reflected signal was demodulated into its in-phase ($I$) and quadrature ($Q$) components (Supplementary Information~\ref{sec:time-domain-measurement}). After the incident pulse is switched off, the detected power $I^2(t)+Q^2(t)$ is proportional to the energy stored in the resonator, $|a(t)|^2$, during the ring-down, where $|a(t)|$ denotes the intracavity field amplitude. The reflected signal is recorded as a function of time and  $V_\mathrm{BO}$, demonstrating clear oscillatory dynamics corresponding to time-domain energy exchange between the plasmon mode and RF photons. These dynamics manifest as Rabi-like oscillations~\cite{Zhang2016-ft,Zhang2014-kc}, i.e., classical normal-mode beating, and are most pronounced near zero-detuning ($V_\mathrm{BO}=-29$~ V in Fig.~\ref{fig:strong_coupling_time_domain}(c)). On resonance, the energy stored in the resonator exhibits a decaying oscillation:
\begin{equation}
|a(t)|^2 \propto
e^{-\frac{\sum_\kappa}{2} t}\,
\cos^2(\Lambda_0 t + \phi), \label{eq:a_square}
\end{equation}
where $\sum_\kappa=\kappa+\gamma_p$ and  $\Lambda_0 $ is the oscillation frequency given by Eq.~\ref{eq:Lambda} (Sec.~\ref{sec:coupled-mode} in Methods). By fitting this expression to the measured $I^2+Q^2$ trace at $V_\mathrm{BO}=-29~\mathrm{V}$ where $\Delta \approx 0$, we extract $g/2\pi = 4.906 \pm 0.009~\mathrm{MHz}$ and $\gamma_{\mathrm p}/2\pi = 3.30 \pm 0.02~\mathrm{MHz}$ (inset of Fig.~\ref{fig:strong_coupling_time_domain}(c)). These values slightly differ from those obtained in the frequency-domain measurements ($g/2\pi = 4.55 \pm 0.02~\mathrm{MHz}$ and $\gamma_\mathrm{p}/2\pi = 5.10 \pm 0.07~\mathrm{MHz}$). This discrepancy may arise because, during the time-domain experiment, no continuous RF power is applied to the device, which can lead to decreases in the electron temperature and thereby modify the effective coupling and dissipation mechanisms (Supplementary Information~\ref{sec:Wigner}). The calculated time-domain response shown in Fig.~\ref{fig:strong_coupling_time_domain}(d) reproduces the main features
of the experimental data. Off resonance, the decay is governed by the resonator decay rate $\kappa$, while on resonance (zero detuning) it is given by $(\kappa+\gamma_\mathrm{p})/2$ due to hybridization with the plasmon mode. Because $\gamma_\mathrm{p}>\kappa$, the decay is slower in the off-resonant regime. At the same time, for finite detuning $\Delta=\omega_0-\omega_\mathrm{p}$ in the off-resonant regime, the oscillation frequency increases to $\Lambda \approx \sqrt{g^2+\left(\frac{\Delta}{2}\right)^2}$, resulting in faster oscillations than on resonance.

\section{Temperature-dependent plasmon decay rate and Wigner crystal transition}\label{sec:temp_dep_gamma_p}

\begin{figure}[h!]
    \centering
    \includegraphics[width=1\linewidth]{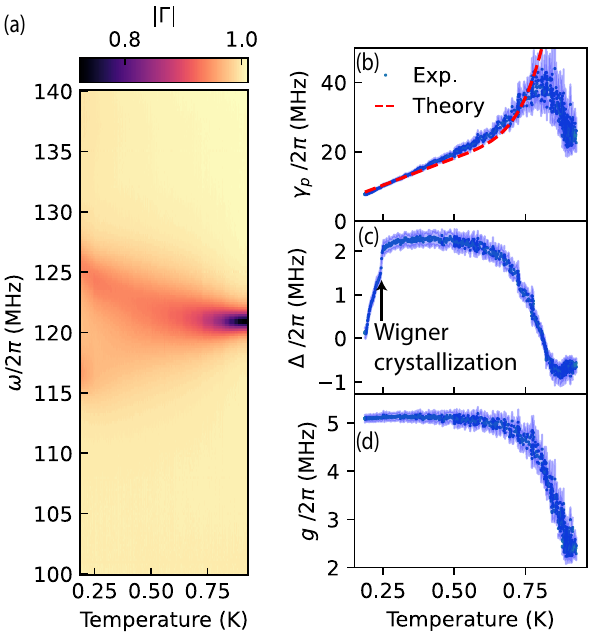}
    \caption{The RF excitation power is set to \(-50~\mathrm{dBm}\) at the output of the signal generator. (a) Resonance peaks measured at \(V_\mathrm{BO} = -29~\mathrm{V}\) as a function of
temperature; other measurement
conditions are identical to those in Fig.~\ref{fig:strong_coupling_time_domain}(a).
(b,c,d) Temperature dependence of the plasmon decay rate
$\gamma_{\mathrm{p}}$ in (b), the detuning $\Delta = \omega_0 - \omega_{\mathrm{p}}$
in (c), and the coupling strength $g$ in (d), obtained from fit of Eq.~\ref{eq:reflection_coefficient_input_output} to the data in (a).
Blue circles represent the extracted values, while the semi-transparent blue lines indicate the
fitting uncertainties (95\% confidence intervals, as used
throughout this manuscript). In (b), the red dashed line shows \(\gamma_{\mathrm{p}} = 1/\tau\),
where \(\tau\) is the momentum--relaxation time calculated using the theoretical
model reported in Ref.~\onlinecite{Saitoh1978-jc} for a pressing field of
\( 92~\mathrm{V/cm}\).
}\label{fig:temp_dep_gamma_p}
\end{figure}

Under the same voltage configurations that satisfies the strong-coupling condition discussed in the previous section, we increased the temperature and observed that the resonance split peaks gradually disappeared (Fig.~\ref{fig:temp_dep_gamma_p} (a)). The disappearance of the split peaks at higher temperatures results from the increase of the plasmon decay rate $\gamma_{\mathrm{p}}$, which violates the strong-coupling condition requiring \(
\Lambda_0
\) to be real. Note that this effect is not caused by electron loss. If the electrons were absent, the resonance peak should be a sharp single peak characterized only by the decay rate of the LC
resonator, $\kappa/2\pi = 0.39~\mathrm{MHz}$. In contrast, at elevated temperatures the linewidth is governed by the much larger plasmon decay rate $\gamma_{\mathrm{p}}$, confirming that the peak merging originates from enhanced plasmon decay.
We fitted the temperature-dependent resonance peaks at each temperature using Eq.~\ref{eq:reflection_coefficient_input_output} and extracted the plasmon decay rate $\gamma_{\mathrm{p}}$ (Fig.~\ref{fig:temp_dep_gamma_p} (b)), the detuning $\Delta=\omega_0 - \omega_{\mathrm{p}}$ (Fig.~\ref{fig:temp_dep_gamma_p} (c)), and the coupling strength $g$ (Fig.~\ref{fig:temp_dep_gamma_p}(d)). 

Plasmon decay is related to electron scattering via $\gamma_{\mathrm{p}} = 1/\tau$, where $\tau$ denotes the electron momentum--relaxation time. For electrons on helium, $\tau$ is governed by electron--ripplon and electron-helium vapor scattering: ripplons are quantized surface waves of liquid helium, and thermally populated long-wavelength ripplon modes dominate below 700~mK for the pressing field used here, while scattering by helium vapor atoms becomes the main relaxation mechanism above 700~mK~\cite{Monarkha2004-un,Kawakami2021}. These processes determine the temperature dependence of $\tau$, and hence of $\gamma_{\mathrm{p}}$. The measured plasmon decay rate $\gamma_{\mathrm{p}}$ agrees well with
theoretical calculations neglecting many-electron effects in the
zero-frequency (DC) limit for an effective pressing field of
$92~\mathrm{V/cm}$~\cite{Saitoh1978-jc}, obtained from $V_\mathrm{BC}/D$ with a correction for image-charge fields due to the electrons being located just below the midplane at $0.48D$~\cite{Isshiki2007-oa} (Fig.~\ref{fig:temp_dep_gamma_p}(b)). This temperature dependence is also consistent with previously reported experimental results
obtained under comparable pressing-field conditions from
low-frequency mobility measurements~\cite{Mehrotra1984-rv, Kirichek1998-xy}
and plasmon absorption measurements~\cite{Kirichek1998-xy}.
At the electron density of this experiment
($n \sim 10^{8}~\mathrm{cm^{-2}}$),
Ref.~\onlinecite{Mehrotra1984-rv} suggested that many-electron effects
begin to emerge in this density regime, although they remain relatively weak~\cite{Ikegami2017-fm,Monarkha2004-un}. The drop of $g$ observed for $T \gtrsim 700~\mathrm{mK}$
(Fig.~\ref{fig:temp_dep_gamma_p}(d)) is likely due to a partial loss of
electrons. In the same temperature range, the apparent reduction of
$\gamma_\mathrm{p}$ can be attributed to a decrease in the number of electrons
participating in the collective plasmon mode, together with an increased
uncertainty in the fitting procedure.

The detuning
\(
\Delta 
\)
shows a pronounced change around \(250~\mathrm{mK}\). This behavior is associated with a rapid change in the plasmon frequency \( \omega_{\mathrm{p}} \) due to the formation of a Wigner crystal. Below this temperature, the electrons are expected to crystallize into the Wigner crystal phase, whereas above it they remain in the liquid phase. For temperature $T \approx 250~\mathrm{mK}$ and  2D electron density of $n_0 \approx 10^{8}~\mathrm{cm^{-2}}$, the 2D plasma parameter (also referred to as the Coulomb coupling parameter) is estimated as
\[
\Gamma = \frac{e^2}{4\pi\varepsilon_0 k_{\mathrm{B}} T}\sqrt{\pi n_0} \approx 120,
\]
which is consistent with previously reported values~\cite{Grimes1979-rv,Andrei1997Two-DimensionalSubstrates,Morf1979-iq,Monarkha2004-un,Ikegami2010-xx}. In this regime, the periodic electron lattice induces a commensurate deformation of the helium surface called the dimple sublattice, leading to stronger coupling between plasmons and ripplons. As a consequence, multiple branches of coupled plasmon-ripplon modes emerge. The highest-frequency branch, referred to as the optical branch, has a frequency
$\sqrt{\omega_{\mathrm{p}}^{2} + \omega_{d}^{2}}$,
where $\omega_d$ denotes the frequency of electrons oscillating in the local dimple potential, which we refer to as the dimple frequency. With decreasing temperature, the dimple potential deepens and $\omega_d$ increases toward its zero-temperature value $\omega_d^0 = 75~\mathrm{MHz}$ for a pressing field of 92~V/cm~\cite{Monarkha2004-un}. As a result, the optical mode frequency becomes higher and the detuning $\Delta$ shifts from its value in the electron fluid phase.  From the observed $\Delta = 2~\mathrm{MHz}$ at our lowest temperature
($167~\mathrm{mK}$), we evaluate the dimple frequency to be
$\omega_d/2\pi = 29~\mathrm{MHz}$, consistent with the above picture.

In contrast, no clear signature of Wigner crystallization is observed in the plasmon decay rate $\gamma_{\mathrm{p}}$. This is consistent with earlier work~\cite{Kirichek1998-xy}, which showed that the plasmon linewidth measured using absorption spectroscopy remains insensitive to Wigner crystallization, whereas the mobility measured at low-frequency ($\lesssim 1~\mathrm{MHz}$) exhibits a rapid decrease. This difference originates from the distinct probe-frequency regimes. At low frequencies, the motion of the Wigner crystal drags the dimple sublattice
along at the same velocity, which effectively increases the electron mass
$m_e$ by the dimple mass $M_d$ ($M_d/m_e \sim$ several hundred). This large mass
enhancement renders the inertia term non-negligible at lower
frequencies ($\lesssim 1~\mathrm{MHz}$), resulting in a significant reduction
of the mobility in the Wigner crystal phase~\cite{Monarkha2012-sg}. In contrast, in the optical branch, the Wigner crystal and the dimple
sublattice oscillate $180^\circ$ out of phase. Because of the large mass
difference between $m_e$ and $M_d$, the dimple sublattice remains nearly
stationary, and the electrons oscillate with only a small amplitude around the
bottoms of the dimples. In this regime, theoretical studies
~\cite{Monarkha2001-ii,Monarkha2004-un} indicate that the scattering rate is nearly
the same in the Wigner crystal and electron liquid phases, and that there is no
significant difference between the zero- and high-frequency scattering rates,
except at very low temperatures ($k_B T < \hbar \omega$), where \(k_\mathrm{B}\) is the Boltzmann constant and \(h\) is Planck’s constant. These predictions are
consistent with our observations in Fig.~4(b).


\section{Conclusion and outlook}\label{sec:concl}

We demonstrate the hybridization of RF electromagnetic fields and plasmons of
electrons on liquid helium. By tuning the DC voltages applied to the electrodes
beneath the electron layer, both the plasmon frequency and the coupling strength
can be controlled. The excellent agreement between experiment and
theoretical calculations highlights the  purity of the
electron-on-helium system, which enables accurate modeling and quantitative
simulation of the coupled dynamics. The LC resonator--based plasmon readout technique also provides a useful means to study electron dynamics on helium such
as the Wigner crystal transition and the temperature-dependent electron scattering process.

The demonstration of strong coupling between plasmons and an LC resonator
reported here constitutes an important step toward using plasmons as carriers of
quantum information. In the present work, the plasmon frequency
\(\omega_\mathrm{p}/2\pi \approx 120~\mathrm{MHz}\) lies below the thermal energy
 \(k_\mathrm{B}T/h\) at the operating temperature of \(T \approx 180~\mathrm{mK}\).
Under these conditions, thermal noise is expected to dominate, rendering
plasmons unsuitable for storing quantum information. A natural next step is therefore to confine plasmons to smaller
spatial regions to increase their frequency into the gigahertz range. Plasmon
modes in the 4--8~GHz range have already been demonstrated for electrons on
helium, matching the operating frequencies of superconducting resonators%
~\cite{Koolstra2019-mq,Mikolas2025-fb}. Achieving strong coupling in this
frequency range would enable operation in the quantum regime, where plasmons can
coherently exchange quantum information with a microwave photon in a
superconducting resonator.

\section{Methods}\label{sec:methods}

\subsection{Simulation of the plasmon frequency and impedance}\label{sec:Green_Sim}

In this section, we describe how the plasmon frequencies (dashed lines in Fig.~\ref{fig2}(c)) and the plasmon impedance $Z_\mathrm{p}$ used to calculate the reflection coefficient $\Gamma$ in Fig.~\ref{fig2}(c) are obtained. The calculation is based on a self-consistent charge-density equation derived from the continuity equation and Coulomb interactions, which is solved using only the simulated electron density profiles as input.
\subsubsection*{Plasmon frequency}
Charge conservation in the 2D electron system is described by
the continuity equation, which relates the time derivative of the charge
density to the current flow. Assuming azimuthal symmetry in cylindrical
coordinates, the continuity equation for the electron sheet reads
\begin{equation}
   -e \frac{\partial  n(r,t)}{\partial t}
    +
    \frac{1}{r}\frac{\partial}{\partial r}
    \left[ r\, j_r(r,t) \right]
    = 0 ,
    \label{eq:continuity_cont}
\end{equation}
where $n(r,t)=n^0(r)+\delta n(r,t)$. Here, $n^0(r)$ is the static electron
density profile, as shown in Fig.~\ref{fig2}(b,d), and $\delta n(r,t)$ denotes
the oscillating electron density fluctuation.
The RF voltage $v e^{-i\omega t}$ applied to the top center electrode
acts as the external excitation, generating an oscillating electric field
$\delta \mathbf{E}^\mathrm{RF}(r,t)= \delta E^\mathrm{RF}_r(r,t) \bm{r} + \delta E^\mathrm{RF}_z(r,t) \bm{z}$ in the electron layer and inducing plasmonic charge motion. Owing to the cylindrical symmetry of the excitation, the electric field
has only $r$ and $z$ components, while the current density is purely radial and
confined to the 2D electron sheet. The radial current density depends linearly on the local electric field and is given by
\begin{equation}
    j_r(r,t)
    =
    \sigma(r)\, \delta E_r(r,t),
\end{equation}
where the 2D conductivity is
\begin{equation}
    \sigma(r)
    =
    \frac{e^2 n^0(r)\tau}{m}
    \frac{1}{1-i\omega\tau}.
    \label{eq:conductivity_cont}
\end{equation}
The oscillating electric field consists of the externally applied RF component
and the self-consistent field generated by electron density fluctuations,
\begin{equation}
    \delta E_r(r,t)
    =
    \delta E_r^{\mathrm{RF}}(r,t)
    +
    \delta E_r^{\mathrm{pl}}(r,t).
\end{equation}

Correspondingly, the radial current density can be decomposed into an externally
driven component and a plasmon-induced component,
\begin{equation}
    j_r(r,t)
    =
    j_r^{\mathrm{RF}}(r,t)
    +
    j_r^{\mathrm{pl}}(r,t),
\end{equation}
with
\begin{align}
    j_r^{\mathrm{RF}}(r,t) &= \sigma(r)\,\delta E_r^{\mathrm{RF}}(r,t), \\
    j_r^{\mathrm{pl}}(r,t) &= \sigma(r)\,\delta E_r^{\mathrm{pl}}(r,t).
    \label{eq:jpl_def}
\end{align}
The plasmon-induced component $\delta E_r^{\mathrm{pl}}$ represents the
self-consistent internal restoring field arising from Coulomb interactions
between electrons~\cite{Wilen1988-ui,Prasad1987-sn}. It is given by
\begin{equation}
    \delta E_r^{\mathrm{pl}}(r,t)
    =
    - \int dr' \,
    \delta n(r',t)\,
    \partial_r G(r,r'),
    \label{eq:Epl_cont}
\end{equation}
where  $\partial_r G(r,r')$ is the radial derivative of the Green’s
function describing the electric field at radius $r$ generated by a unit charge
fluctuation at $r'$. This internal field acts as a restoring field: any spatial
modulation of the electron density produces a charge imbalance that generates
an electric field tending to suppress the modulation, giving rise to collective
plasmon oscillations. Taking $\delta n(r,t)=\delta n(r)e^{-i\omega t}$
(and similarly for fields and currents), Eq.~\eqref{eq:continuity_cont} becomes
\begin{equation}
    i\omega e\,\delta n(r)
    +
    \frac{1}{r}\frac{\partial}{\partial r}
    \left[
        r\,\sigma(r)\left(
            \delta E_r^{\mathrm{RF}}(r) + \delta E_r^{\mathrm{pl}}(r)
        \right)
    \right]
    =0.
    \label{eq:cont_freq_cont}
\end{equation}
Substituting the plasmon-induced electric field
(Eq.~\eqref{eq:Epl_cont}) into the frequency-domain continuity equation
(Eq.~\eqref{eq:cont_freq_cont}) yields an integral equation for the oscillating
electron density $\delta n(r)$,
\begin{equation}
    \int dr'\, \mathcal{H}(r,r';\omega)\,\delta n(r')
    =
    \frac{1}{r}\frac{\partial}{\partial r}
    \!\left[r\,\sigma(r)\,\delta E_r^{\mathrm{RF}}(r)\right],
    \label{eq:H_operator_cont}
\end{equation}
where $\mathcal{H}(r,r';\omega)$ is a linear kernel determined by the device
geometry through the Green’s function $G(r,r')$, the static electron density
profile $n^0(r)$ through the conductivity $\sigma(r)$, and the angular
frequency $\omega$. In the absence of external RF driving,
$\delta E_r^{\mathrm{RF}}(r)=0$, Eq.~\eqref{eq:H_operator_cont} reduces to the
homogeneous integral equation
\begin{equation}
    \int dr'\, \mathcal{H}(r,r';\omega)\,\delta n(r') = 0 .
    \label{eq:H_hom_cont}
\end{equation}
Nontrivial solutions of Eq.~\eqref{eq:H_hom_cont} exist only for discrete values
of $\omega$, which define the plasmon frequencies~\cite{Prasad1987-sn}. Each such
eigenfrequency is associated with an eigenfunction $\delta n(r)$ describing a
standing-wave charge-density oscillation in the radial direction. These
solutions correspond to the $\mu$-th radial plasmon modes,
$\mu = 1, 2, 3, \ldots$.
\subsubsection*{Plasmon impedance}
When an AC voltage $v e^{-i\omega t}$ is applied to the top center electrode,
it generates a radial RF electric field
$\delta E_r^{\mathrm{RF}}(r)$.
According to Eq.~(\ref{eq:H_operator_cont}), this RF field modulates the electron
density, resulting in a density variation $\delta n(r)$.
This density modulation induces a charge variation $\delta Q$ on the top center
electrode~\cite{Wilen1988-ui}. The coupling between the LC resonator and the plasmon mode arises from this
density-induced charge on the top center electrode.
The resulting plasmon impedance is given by~\cite{note:minus_sign}
\begin{equation}
    Z_p = \frac{v}{-i\,\omega\,(\alpha\,\delta Q)}.
    \label{eq:Zp}
\end{equation}
Here, we introduce a factor $\alpha$ to account for the real device, in which a fraction of the induced image charge is shunted into parasitic capacitances, mainly originating from metal plates on the back side of the PCB that are not included in the simulation. We define a participation ratio $\alpha = C_{\mathrm{sim}}/C$, where $C_{\mathrm{sim}}$ is the capacitance obtained from the simulated geometry and $C$ is the capacitance of the real device including parasitic contributions. The latter is extracted as $C = 2.131~\mathrm{pF}$ by fitting Eq.~\ref{eq:Gamma_quality_factor} to the measured resonance peak, using an independently measured inductance $L$~\cite{Jennings2025-ml}. With $C_{\mathrm{sim}} = 0.5~\mathrm{pF}$, we obtain $\alpha \approx 0.25$. As a result, the effective charge contributing to the plasmon response is reduced from $\delta Q$ to $\alpha\delta Q$, as reflected in Eq.~(\ref{eq:Zp}).

In Fig.~\ref{fig2}(a,c), the experimentally extracted coupling strengths are slightly larger than the simulated values, which may reflect minor geometric features of the actual device—such as conducting support structures and exposed $0.1\,\mathrm{mm}$ dielectric regions between electrodes—that are not fully captured in the simulation.

The continuous formulation described above is implemented numerically by
discretizing the system in cylindrical coordinates $(r,z)$. For the simulations corresponding to Fig.~\ref{fig2} and
Fig.~\ref{fig:strong_coupling_time_domain}, the radial coordinate is
discretized over $0 \le r \le 7.5~\mathrm{mm}$ and the vertical coordinate is
discretized over $-1~\mathrm{mm} \le z \le 1~\mathrm{mm}$. The radial direction is
divided into 500 grid points, while the $z$ direction is divided into 200 grid points. All numerical calculations are performed in Python. The codes can be found in Ref.~\onlinecite{plasmon_repo}. The only input parameters of the numerical simulation are the static electron density
profile $n^0(r)$ and the electron momentum-relaxation time $\tau$; no additional
free parameters are introduced.

\subsection{Time-domain measurement \label{sec:time-domain-measurement}}

A time-domain microwave reflectometry measurement was performed to probe the plasmon dynamics. Two microwave sources (Keysight E8267D and Rohde \& Schwarz SMB100A) were phase-locked to a common 10~MHz reference provided by a Stanford Research Systems FS725 rubidium frequency standard. Microwave pulses with a duration of 20~ns were generated using a Tektronix AWG~5204. A Kaiser--Bessel envelope with $\beta=14$ was employed to suppress spectral leakage and unwanted frequency components. The pulse waveform was fed into the I port of the internal IQ modulation mode of a Keysight E8267D, generating RF pulses with a source power of
$-20~\mathrm{dBm}$. Before reaching port~1 of the sample, the signal was
attenuated by a total of $38~\mathrm{dB}$ in the input line inside the
refrigerator, including all inserted attenuators. The reflected signal from port~2 of the sample was amplified first at 4~K by +40 dB using a Cosmic Microwave Technology CITLF3 cryogenic amplifier, further amplified at room temperature  by 20~dB using a Mini-Circuits ZFL-500LN+ low-noise amplifier.  The signal subsequently demodulated using a Polyphase Quadrature Demodulator (AD0105B), with the Rohde \& Schwarz SMB100A serving as the local oscillator reference. The in-phase ($I$) and quadrature ($Q$) voltage components were passed through a Mini-Circuits SLP-50+ low-pass filter and recorded by a Teledyne LeCroy WaveRunner~9054 oscilloscope (500 MHz bandwidth), with 200 waveform averages performed in real time. To remove systematic distortion of the demodulator response (including $I$/$Q$ gain imbalance, phase offsets, and DC offsets), a full complex calibration of the demodulation chain was implemented in software.

The detected power satisfies
\[
I^2(t) + Q^2(t) \;\propto\; |a_{\mathrm{out}}(t)|^2 ,
\]
where \(a_{\mathrm{out}}(t)\) is the output field of the resonator. The resonator input--output relation is
\[
a_{\mathrm{out}}(t)
= a_{\mathrm{in}}(t) - \sqrt{\kappa_{\mathrm{ext}}}\, a(t),
\]
where \(a_{\mathrm{in}}(t)\) denotes the incident field at the resonator. After the incident field is switched off, \(a_{\mathrm{in}}(t) \approx 0\), so that
\[
a_{\mathrm{out}}(t) \approx -\sqrt{\kappa_{\mathrm{ext}}}\, a(t),
\qquad
|a_{\mathrm{out}}(t)|^2 \propto |a(t)|^2 .
\]
Thus, in the ring-down regime, the measured IQ power \(I^2(t) + Q^2(t)\) is directly proportional to the energy stored in the resonator, i.e., the intracavity photon number \( |a(t)|^2 \).

\subsection{Time-domain coupled-mode}\label{sec:coupled-mode}

The equations of motion for the coupled LC resonator mode $a(t)$ 
and the plasmon mode $b(t)$ are
\begin{equation}
\frac{d}{dt}
\begin{pmatrix}
a \\
b
\end{pmatrix}
=
\begin{pmatrix}
-\kappa/2 - i\omega_0  &  -i g \\
-i g                     &  -\gamma_p/2 - i\omega_p
\end{pmatrix}
\begin{pmatrix}
a \\
b
\end{pmatrix}, \label{eq:EqOfMotion}
\end{equation}
where $\kappa$ and $\gamma_p$ denote the decay rates of the two modes. The eigenvalues of this dynamical matrix are $-i \tilde{\omega}_\pm$ where
\begin{equation}
 \tilde{\omega}_\pm=\omega_0 - \frac{\Delta}{2} - i \frac{\Sigma_\kappa}{4} \pm \Lambda.
    \label{eq:omega_pm_tilde}
\end{equation}
where 
\begin{equation}
     \Lambda=\sqrt{\left(\frac{\Delta}{2} - i \frac{\Delta_\kappa }{4} \right)^2 +g^2},
     \label{eq:Lambda_general}
\end{equation}
$\Sigma_\kappa=  \kappa + \gamma_p$, $\Delta_\kappa= \kappa - \gamma_p$, and the detuning is defined as $\Delta = \omega_0 - \omega_p $. For the initial condition $a(0)=1$ and $b(0)=0$, the mode amplitudes can be
written as a superposition of the two normal modes,
\begin{equation}
a(t)
=
A\,e^{-i\tilde{\omega}_+ t}
+
(1-A)\,e^{-i\tilde{\omega}_- t},
\end{equation}
\begin{equation}
b(t)
=
\alpha_+ A\,e^{-i\tilde{\omega}_+ t}
+
\alpha_-(1-A)\,e^{-i\tilde{\omega}_- t},
\end{equation}
with
$\alpha_\pm=\frac{- \frac{\Delta}{2} + i \frac{\Delta_\kappa}{4} \pm \Lambda}{g}$
and
$A=\frac{1}{2} \left( 1+ \frac{ \frac{\Delta}{2} - i \frac{\Delta_\kappa}{4} }{\Lambda} \right)$. The resulting cavity-field amplitude is therefore given by
\begin{equation}
|a(t)|
=
e^{-  \frac{\Sigma_\kappa}{4} t }
\left|
A\,e^{-i\Lambda t}
+
(1-A)\,e^{i \Lambda t}
\right|,
\label{eq:a_general}
\end{equation}
which describes oscillatory energy exchange with an overall exponential decay. On resonance ($\Delta=0$), this expression simplifies to
\begin{equation}
|a(t)|
=
e^{-\frac{\Sigma_\kappa}{4}t}\,
\left|
\cos(\Lambda_0 t)
-
\frac{\Delta_\kappa}{4\Lambda_0}\,
\sin(\Lambda_0 t)
\right|,
\end{equation}
with $\Lambda_0=\sqrt{ g^2-\left( \frac{\Delta_\kappa }{4} \right)^2}$. The intra cavity photon number, proportional to $|a(t)|^{2}$, is then given by
\begin{equation}
|a(t)|^{2}
=
e^{- \frac{\Sigma_\kappa}{2} t }\,
\left(1+\left(\frac{\Delta_\kappa}{4\Lambda_0} \right)^{2}\right)
\cos^{2}(\Lambda_0 t + \phi),
\label{eq:a_square_general}
\end{equation}
with $\phi=\arctan\left ( \frac{\Delta_\kappa}{4\Lambda_0} \right)$. This expression describes an exponentially decaying, Rabi-like exchange
of energy between the cavity and plasmon modes, in agreement with the time-resolved response shown in the inset of Fig.~\ref{fig:strong_coupling_time_domain}(c).

\vspace{10pt}

\begin{acknowledgments}
This work was supported by the RIKEN Hakubi Program, the RIKEN Center for Quantum Computing, JST-FOREST, the Hattori Hokokai Foundation, the National Natural Science Foundation of China (Grant No. 12474135), and the National Basic Research Program of China (Grant No. 2025YFA1411400). We are grateful to Prof. Jaw-Shen Tsai for granting us access to the Kelvinox 400HA dilution refrigerator. We thank Prof. Denis Konstantinov for useful discussions.
\end{acknowledgments}

\section*{Author contributions} 
A.J. led the experiments and data analysis.
Y.T. assisted with the experiments.
I.G. and O.R. contributed to technical development and experimental implementation.
I.J.B., T.G., H.I., A.J., and E.K. performed theoretical modeling and numerical simulations, with contributions from J.W.
E.K. conceived and supervised the project.
The manuscript was written by A.J., H.I. and E.K. with input from all authors.
\appendix

\section{Approximated plasmon frequency calculation} \label{sec:plasmon_freq}

To approximately obtain the plasmon frequency from Eq.~\ref{eq:omega_p}, we extract an effective radius $R^*$ and an effective uniform density $n_0$ from
the simulated profile following Ref.~\onlinecite{Glattli1985-fp}. 
The wave number of the \(\mu\)-th radial mode is determined by the boundary
condition that the radial current vanishes at \(R^*\), which is given by
\(
J_0'(k_{0,\mu} R^*) = 0,
\)
where \(J_0'\) denotes the derivative of the zeroth-order Bessel function. We take $R^*$ as the radius at which the simulated density
drops to zero. With this \(R^*\), the uniform electron density \(n_0\) is obtained by
approximating the density profile as a disk of radius \(R^*\), whose total
electron number matches that obtained from the simulated density profile in
Fig.~\ref{fig2}(b,d).

  Fig.~\ref{fig2}(d) shows the simulated electron density profiles corresponding
to the resonances observed at
$V_{\mathrm{BC}} = 10, 9, 8,$ and $7.5~\mathrm{V}$.
For each value of $V_{\mathrm{BC}}$, we select the resonance appearing at the
most negative value of $V_{\mathrm{BM}}$
($V_{\mathrm{BM}}^\mathrm{sim} = -14.2, -17.9, -23.1,$ and $-26.4~\mathrm{V}$, respectively).
The corresponding pairs of $(k_{0,1}, n_0)$ for each density profile are listed
in Table~\ref{tab:plasmon_modes}. 
Substituting these values into Eq.~\ref{eq:omega_p} yields
$\omega_{\mathrm{p}}/2\pi \sim 125~\mathrm{MHz}$ in all cases, indicating that these
resonances correspond to the fundamental radial plasmon mode. The deviation from $\omega_0/2\pi$ reflects the use of an effective uniform-density model. The thick red line in Fig.~\ref{fig2}(b) corresponds to the density
profile for
$V_{\mathrm{BC}} = 7.5~\mathrm{V}$ at $V_\mathrm{BM}^\mathrm{exp}= 0.8~\mathrm{V}$.  
In this case, substituting $(k_{0,2}, n_0)$ (Table~\ref{tab:plasmon_modes}) into Eq.~\ref{eq:omega_p} gives $\omega_{\mathrm{p}}/2\pi =130.5~\mathrm{MHz}$, demonstrating that this resonance corresponds to the second radial plasmon mode.

\begin{table}[H]
    \centering
    \caption{Extracted parameters $(R^{*}, k_{0,\mu}, n_{0})$ for representative resonances
using the simulated density profiles (see, for example Fig.~\ref{fig2}(b,d)) for the $V_\mathrm{BM}^\mathrm{exp}$ at which avoided level crossings in Fig.~\ref{fig2}(a)  and the corresponding $\mu$-th plasmon frequencies $\omega_{\mathrm{p}}$ calculated from Eq.~\ref{eq:omega_p}. The corresponding simulated values $V_{\mathrm{BM}}^{\mathrm{sim}}$, at which
avoided level crossings occur in the simulation (Fig.~2(c)), are also listed
for comparison.
 OOR indicates the plasmon mode was outside the simulated $V_\mathrm{BM}$ range.}
    \label{tab:plasmon_modes}
    \begin{tabular}{cccccccc}
        \hline\hline
        $V_{\mathrm{BC}}$ &
        $\mu$ &
        $V_{\mathrm{BM}}^{\mathrm{exp}}$ &
        $V_{\mathrm{BM}}^{\mathrm{sim}}$ &
        $R^\ast$ &
        $k_{0,\mu}$ &
        $n_0$ &
        $\omega_{\mathrm{p}}/2\pi$\\
        (V) & & (V) & (V) & ($\mathrm{cm}$)& ($\mathrm{cm}^{-1}$) & ($10^7\mathrm{cm}^{-2}$) & (MHz)\\
        \hline
        10   & 1 & -14.2 &  -13.8  & 0.3270 & 11.72 & 4.14 & 126.9 \\
        9   & 1 & -17.9 &  -18.4  & 0.3150 & 12.16 & 3.81 & 125.0 \\
        8   & 1 & -23.1 &  -24.0  & 0.2985 & 12.84 & 3.50 & 124.5 \\
        7.5   & 1 & -26.4 &  -27.0  & 0.2910 & 13.17 & 3.41 & 125.0 \\
        \hline
        10   & 2 & 4.0 &  OOR  & 0.3870 & 18.13 & 2.92 & 142.0 \\
        9   & 2 & 2.9 &  OOR  & 0.3780 & 18.56 & 2.65 & 137.3 \\
        8   & 2 & 1.5 &  3.8  & 0.3660 & 19.17 & 2.36 & 132.0 \\
        7.5   & 2 & 0.8 &  2.6  & 0.3600 & 19.49 & 2.26 & 130.5 \\
        6.0   & 2 & -3.3 &  -2.2  & 0.3300 & 21.26 & 1.93 & 126.7 \\
        4.0   & 2 & -14.3 &  -13.8  & 0.2745 & 25.56 & 1.51 & 123.9 \\
        \hline
        6.0   & 3 & 1.9 &  4.6  & 0.3600 & 28.26 & 1.64 & 135.8 \\
        4.0   & 3 & -3.5 &  -2.4  & 0.3090 & 32.92 & 1.20 & 125.9 \\
        \hline
        6.0   & 4 & 3.5 &  OOR  & 0.3735 & 35.67 & 1.53 & 147.8 \\
        4.0   & 4 & 0.5 &  1.8  & 0.3330 & 40.01 & 1.03 & 128.5 \\
        \hline\hline
    \end{tabular}
\end{table}

\section{Reflection coefficient } \label{sec:ref_coeff}

The reflection coefficient of the LC resonator without plasmon (Eq.~\ref{eq:Ref_coeff} with Eq.~\ref{eq:Z_l_delta_omega} in the limit $Z_p \to \infty$) can be approximated as 
\begin{equation}
    \Gamma_\mathrm{ref}
    \approx
    1 - \frac{2 Q_\mathrm{tot}/Q_\mathrm{ext}}{1 + i\,2 Q_\mathrm{tot}
    \left( \frac{\omega}{\omega_0} - 1 \right)}=1 - \frac{-i\,\kappa_\mathrm{ext}}
    {(\omega - \omega_0) - i\,\kappa/2}
    \label{eq:Gamma_quality_factor}
\end{equation}
for $C_c \ll C$. The resonator resonance frequency and the quality factors are given by
\begin{align}
    \omega_0
    &= \sqrt{\frac{1}{(C + C_c)L}}, \\
    Q_\mathrm{int}^{-1}
    &= \frac{1}{R \omega_0 C}
     = \frac{\kappa_\mathrm{int}}{\omega_0}, \\
    Q_\mathrm{ext}^{-1}
    &= \frac{\omega_0 C_c^2 Z_0}{C}
     = \frac{\kappa_\mathrm{ext}}{\omega_0}, \\
    Q_\mathrm{tot}^{-1}
    &= \frac{1}{\omega_0 C}
       \left(
           \frac{1}{R}
           +
           \frac{(\omega_0 C_c Z_0)^2}{Z_0}
       \right)
     = \frac{\kappa}{\omega_0}.
\end{align}
The plasmon mode can be modeled as a linear harmonic oscillator characterized
by a resonance frequency $\omega_{\mathrm{p}}$, a linewidth
$\gamma_{\mathrm{p}}$, and a coupling strength $g$ to the resonator. As a
result, its contribution to the circuit response appears as a
frequency-dependent self-energy term,
\(\frac{g^2}{(\omega - \omega_{\mathrm{p}}) + i\,\gamma_{\mathrm{p}}/2},
\) which, when incorporated into the resonator response, yields the reflection coefficient including the plasmon contribution
(Eq.~\ref{eq:reflection_coefficient_input_output}).

\subsection*{Formula for the Split Resonance Peaks}
We can rewrite Eq.~\ref{eq:reflection_coefficient_input_output} as $ \Gamma_\mathrm{ref}=\frac{A-iB}{A-iC}$, where 
\begin{align}
    A&=(\omega_0-\omega)-\frac{g^2(\omega_\mathrm{p}-\omega)}{(\omega_\mathrm{p}-\omega)^2+(\gamma_\mathrm{p}/2)^2}\\
    B&=\frac{\kappa_\mathrm{int}-\kappa_\mathrm{ext}}{2}+ \frac{g^2 \gamma_\mathrm{p}/2}{(\omega_\mathrm{p}-\omega)^2+(\gamma_\mathrm{p}/2)^2} \\
    C&=\frac{\kappa_\mathrm{int}+\kappa_\mathrm{ext}}{2}+ \frac{g^2 \gamma_\mathrm{p}/2}{(\omega_\mathrm{p}-\omega)^2+(\gamma_\mathrm{p}/2)^2}.
\end{align}
When the denominator of $\Gamma_{\mathrm{ref}}$ vanishes, the system supports non-trivial solutions corresponding to the upper and lower hybridized eigenmodes, yielding the complex eigenfrequencies $\tilde{\omega}_{\pm}$ given in Eq.~\ref{eq:omega_pm_tilde}. At zero detuning, the amplitude of $\Gamma_{\mathrm{ref}}$ exhibits a pair of split resonance peaks when the mode-splitting parameter \(
\Lambda_0 = \sqrt{\, g^{2} - \left( \frac{\Delta_\kappa}{4} \right)^{2} }
\) is real. In this regime, the frequency separation between the two
hybridized modes is $2\Lambda_0$.

We fitted the resonance peaks in Fig.~\ref{fig2}(a), Fig.~\ref{fig:strong_coupling_time_domain}(a) and Fig.~\ref{fig:temp_dep_gamma_p}(a) using the expression
\begin{equation}
    |\Gamma_\mathrm{ref}| = \sqrt{\frac{A^2 + B^2}{A^2 + C^2}}, \label{eq:Gamma_ref}
\end{equation}
from which we extracted the coupling strength \( g \) and the plasmon decay rate \( \gamma_\mathrm{p} \). 

\begin{figure}[t!]
    \centering
    \includegraphics[width=0.9\linewidth]{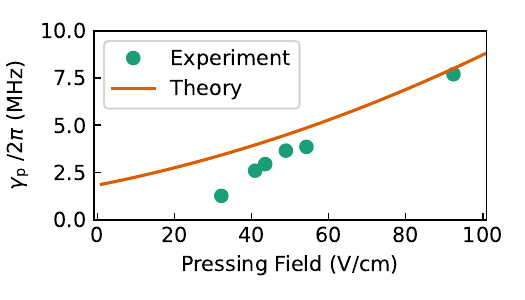}
    \caption{Extracted values of $\gamma_\mathrm{p}$/$2\pi$ as a function of pressing field measured at $T = 180~\mathrm{mK}$. The pressing field corresponds to $V_\mathrm{BC}$/$D$ (for $V_\mathrm{BC} = 10,\ 9,\ 8,\ 7.5,\ 6~\mathrm{V}$) plus a correction for the field due to the image charge induced by the electrons, which is the electric field experienced by the electrons near the center. The value at $V_\mathrm{BC} = 17~\mathrm{V}$ is extracted from Fig.~\ref{fig:strong_coupling_time_domain}(a), while the values at $V_\mathrm{BC} = 10,\ 9,\ 8,\ 7.5,\ 6,\ 4~\mathrm{V}$ are extracted from Fig.~\ref{fig2}(a).
    The value of $\gamma_\mathrm{p}$ for $V_\mathrm{BC}=4~\mathrm{V}$ in Fig.~\ref{fig2}(a) could not be extracted~\ref{fig2}(a) as the plasmon frequency changes rapidly with the radius at higher modes. The orange line indicates $\gamma_{\mathrm{p}} = 1/\tau$ from the model of
Ref.~\onlinecite{Saitoh1978-jc} at $T = 180~\mathrm{mK}$. The marker size is large than the error.}
    \label{fig:Fig_gamma_field}
\end{figure}

The pressing-field dependence of $\gamma_\mathrm{p}$ extracted from Fig.~\ref{fig2}(a) and Fig.~\ref{fig:strong_coupling_time_domain}(a) is presented in Fig.~\ref{fig:Fig_gamma_field}. The extracted values for $V_\mathrm{BC} = 10,\ 9,\ 8,\ 7.5,\ 6~\mathrm{V}$ from Fig.~\ref{fig2}(a) are
\(
\gamma_\mathrm{p}/2\pi = 3.86 \pm 0.07,\ 3.66 \pm 0.04,\ 2.96 \pm 0.02,\ 2.60 \pm 0.04,\ 1.26 \pm 0.06~\mathrm{MHz},
\)
and are used in the simulation shown in Fig.~\ref{fig2}(c). The value of $\gamma_\mathrm{p}$ at $V_\mathrm{BC} = 4~\mathrm{V}$ in Fig.~\ref{fig2}(a) could not be extracted because the plasmon frequency varies rapidly with radius for higher-order modes. Therefore, the value obtained at $V_\mathrm{BC} = 6~\mathrm{V}$ was used in the simulations for $V_\mathrm{BC} = 4~\mathrm{V}$ in Fig.~\ref{fig2}(c).

The temperature dependence of  \( \gamma_\mathrm{p} \), $\Delta=\omega_0-\omega_\mathrm{p}$, and \( g \) extracted from Fig.~\ref{fig:temp_dep_gamma_p} (a) are shown in Fig.~\ref{fig:temp_dep_gamma_p} (b-d). The values \( \kappa_\mathrm{ext}/2\pi = 0.20~\mathrm{MHz} \) and \( \kappa_\mathrm{int}/2\pi = 0.19~\mathrm{MHz} \) used in the fits were obtained from the plasmon-unperturbed (far-from-resonance) configuration~\cite{Jennings2025-ml}.

\section{Coupling strength $g$}\label{sec:coupling_g}

\begin{figure}[h!]
    \centering
    \includegraphics[width=0.9\linewidth]{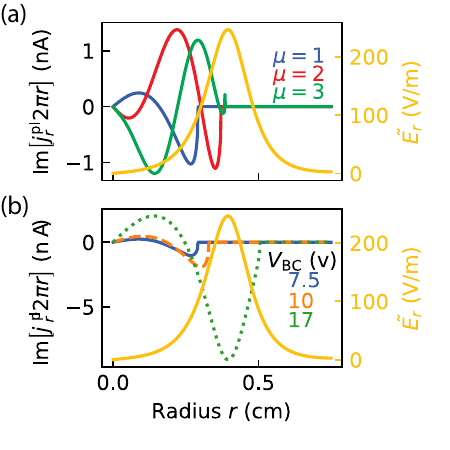}
 \caption{ Radial profiles of the product of the imaginary part of the plasmon current density and the circumference, $\mathrm{Im}[j_r^\mathrm{pl} 2\pi r]$, as a function of the radial coordinate $r$, together with the RF-induced electric field $\delta E_r^\mathrm{RF}$, both calculated for a 1~V AC excitation applied to the top center electrode.
The solid blue, red, green, and orange dashed lines are evaluated at the resonance conditions shown in Fig.~\ref{fig2}(c), as detailed below:
the blue solid lines in (a) and (b) correspond to the fundamental mode ($\mu = 1$) at $V_\mathrm{BC} = 7.5~\mathrm{V}$ and
$V_\mathrm{BM} = -27~\mathrm{V}$;
the red line in (a) corresponds to the second mode ($\mu = 2$) at $V_\mathrm{BC} = 7.5~\mathrm{V}$ and
$V_\mathrm{BM} = 2.6~\mathrm{V}$, and the green line in (a) corresponds to the third mode ($\mu = 3$) at
$V_\mathrm{BC} = 6~\mathrm{V}$ and $V_\mathrm{BM} = 4.6~\mathrm{V}$;
the orange dashed line in (b) corresponds to the fundamental mode ($\mu = 1$) at
$V_\mathrm{BC} = 10~\mathrm{V}$ and $V_\mathrm{BM} = -13.8~\mathrm{V}$.
The green dotted line in (b) corresponds to the fundamental mode ($\mu = 1$) shown in
Fig.~\ref{fig:strong_coupling_time_domain}(b), where $V_\mathrm{BC} = V_\mathrm{BM} = 17~\mathrm{V}$ and strong coupling is observed at
$V_\mathrm{BO} = -30.7~\mathrm{V}$.
}

    \label{fig:Fig_jA_Erf}
\end{figure}

The coupling strength $g$ between the LC resonator field and a plasmon
eigenmode can be quantified by the reactive (imaginary) component of the
power overlap between the plasmon-mode current density and the driving RF
electric field,
\begin{equation}
    g
    \propto
    \mathrm{Im}
    \!\left[
        \int
        \delta E_r^{\mathrm{RF}}(r)\,
        j_r^{\mathrm{pl}}(r)\,
        2\pi r \, dr  \right].
    \label{eq:Cmu_continuous}
\end{equation}
Here, $j_r^{\mathrm{pl}}$ denotes the sheet current density of the plasmon mode, and $\delta E_r^{\mathrm{RF}}$ is the radial RF electric field generated by the RF excitation. Fig.~\ref{fig:Fig_jA_Erf} shows the radial profiles of $j_r^{\mathrm{pl}}(r),2\pi r$ and $\delta E_r^{\mathrm{RF}}(r)$, both calculated for a unit AC excitation voltage. $\delta E_r^{\mathrm{RF}}$ reaches its maximum near
$r \simeq 4~\mathrm{mm}$, corresponding to the edge of the top center
electrode. We numerically verified the validity of the linear relation in Eq.~\ref{eq:Cmu_continuous} by comparing the
values of \( g \) extracted from the resonance splitting of the numerically
simulated reflection spectra shown in Fig.~\ref{fig2}(c) and
Fig.~\ref{fig:strong_coupling_time_domain}(b) with the quantity
\(\mathrm{Im}\!\left[
\int \delta E_r^{\mathrm{RF}}(r)\, j_r^{\mathrm{pl}}(r)\, 2\pi r \, dr
\right]\). A strong enhancement of the coupling strength $g$ occurs when $j_r^{\mathrm{pl}}(r),2\pi r$ closely matches the spatial profile of $\delta E_r^{\mathrm{RF}}(r)$, yielding a large overlap between the two fields, as demonstrated by the green dotted line in Fig.~\ref{fig:Fig_jA_Erf}(b), where strong coupling is observed.

\section*{Derivation of Eq.~\ref{eq:Cmu_continuous}}
We model the system as two coupled harmonic oscillators. The resonator mode is
characterized by an amplitude $A_c(t)$ and an effective mass $M_c$, with kinetic
and potential energies given by
\begin{equation}
K_c = \frac{1}{2} M_c \dot{A}_c^2, \qquad
U_c = \frac{1}{2} M_c \omega_0^2 A_c^2 .
\end{equation}
Similarly, the plasmon mode is described by an amplitude $A_p(t)$ and an
effective mass $M_p$, with kinetic and potential energies
\begin{equation}
K_p = \frac{1}{2} M_p \dot{A}_p^2, \qquad
U_p = \frac{1}{2} M_p \omega_p^2 A_p^2 .
\end{equation}

Coupling between the resonator and plasmon modes is introduced through an
interaction term
\begin{equation}
U_{\mathrm{int}} = G A_c A_p ,
\end{equation}
where $G$ quantifies the coupling strength. The corresponding Lagrangian is
\begin{equation}
L
=
\frac{1}{2} M_c \dot A_c^2
-
\frac{1}{2} M_c \omega_0^2 A_c^2
+
\frac{1}{2} M_p \dot A_p^2
-
\frac{1}{2} M_p \omega_p^2 A_p^2
-
G A_c A_p .
\end{equation}
The Euler--Lagrange equations for $A_c(t)$ and $A_p(t)$ then read
\begin{align}
M_c \ddot A_c + M_c \omega_0^2 A_c + G A_p &= 0 , \\
M_p \ddot A_p + M_p \omega_p^2 A_p + G A_c &= 0 .
\end{align}
Intrpducing a harmonic time dependence
$A_{c,p}(t) \propto e^{-i\omega t}$,
the equations of motion reduce to
\begin{align}
(\omega_0^2 - \omega^2) A_c + G A_p/M_c &= 0 ,\label{eq:Euler1} \\
(\omega_p^2 - \omega^2) A_p +  G A_c/M_p &= 0  \label{eq:Euler2}.
\end{align}
Near resonance ($\omega \approx \omega_0$ and $\omega \approx \omega_p$),   taking 
\begin{equation}
a \equiv \sqrt{\omega_0 M_c}\, A_c , \qquad
b \equiv \sqrt{\omega_p M_p}\, A_p ,
\end{equation}
and
\begin{equation}
g \equiv \frac{G}{2\sqrt{\omega_0 \omega_p M_c M_p}} ,
\end{equation}
Eq.~\ref{eq:Euler1} and Eq.~\ref{eq:Euler2} are reduced to  
\begin{align}
(\omega_0 - \omega) a + gb &= 0 , \\
(\omega_p- \omega)b+  ga &= 0 ,
\end{align}
which is identical to Eq.~\ref{eq:EqOfMotion} for $\gamma_\mathrm{p}=\kappa=0$.

The physical origin of the interaction can be understood in terms of
electromagnetic work. In classical electrodynamics, the instantaneous power
transferred from an electric field $\mathbf{E}$ to a current density
$\mathbf{j}$ is given by $\dot U = \int \mathbf{E} \cdot \mathbf{j}\, dA$.
For time-harmonic fields, only the reactive (non-dissipative) component
contributes to coherent energy exchange between two lossless modes.
Consequently, the coupling constant is determined by the imaginary part of
the field--current overlap,
\begin{equation}
GA_cA_p
=
\frac{1}{\omega}
\mathrm{Im} \int \delta \mathbf{E}_{\mathrm{RF}} \cdot \mathbf{j}_{\mathrm{pl}}\, dA ,
\end{equation}
where the factor $1/\omega$ converts power into energy. Near resonance ($\omega_0 \approx \omega_p$), the coupling rate can finally be written as
\begin{equation}
g
=
\frac{
\mathrm{Im} \int \delta \mathbf{E}_{\mathrm{RF}} \cdot \mathbf{j}_{\mathrm{pl}}\, dA
}{
4\sqrt{U_c U_p}
}.
\end{equation}

\section{Effects of Wigner crystallization and RF Drive on the Plasmon Frequency and Decay}
\label{sec:Wigner}

The RF excitation power differs between the measurements shown in 
Fig.~\ref{fig2} and Fig.~\ref{fig:temp_dep_gamma_p} ($-50~\mathrm{dBm}$) and 
Fig.~\ref{fig:strong_coupling_time_domain} ($-20~\mathrm{dBm}$), both specified at the output of the
signal generator. The higher drive power is required in the time-domain
experiment to excite the plasmon mode strongly enough for real-time observation.

To examine the influence of the RF drive on the resonance condition, we measured
the same voltage configuration as in Fig.~\ref{fig:strong_coupling_time_domain}(a),
but with a reduced RF power of $-50~\mathrm{dBm}$ (data not shown). Under these
conditions, the resonance appears at $V_\mathrm{BO} \approx -25~\mathrm{V}$, instead of
$-29~\mathrm{V}$ for the $-20~\mathrm{dBm}$ drive. As discussed in
Sec.~\ref{sec:temp_dep_gamma_p}, at the density $n_0=10^8~\mathrm{cm}^{-2}$ and at $T \approx 180~\mathrm{mK}$ the electron ensemble
is expected to form a Wigner crystal. We attribute the shift of the resonance to more negative voltages observed at high RF power to approaching to the melting temperature due to heating. The simulated resonance position is found at $V_\mathrm{BO} = -30.7~\mathrm{V}$, slightly different from the
low-power experimental value. This deviation may be similarly due to the heating. Further discussion is provided in
Ref.~\onlinecite{jennings_inprep_2026}.

We also note that the measurements in Fig.~\ref{fig2}
correspond to lower electron densities, where the system remains in the liquid phase.

\section{Temperature dependence of the LC resonator}\label{sec:TempLC}

\begin{figure}[H]
    \centering
    \includegraphics[width=0.9\linewidth]{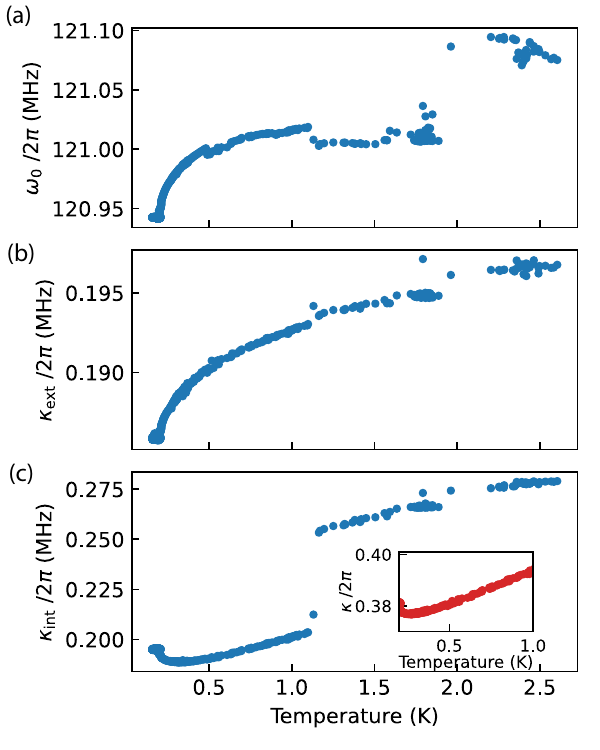}
    \caption{The fitted bare LC resonator parameters as a function of the cell temperature for (a) the resonance frequency, (b) the external decay and (c) the internal decay. The large, sudden increase in \(\kappa_\mathrm{int}\) is attributed to the superconducting transition of the aluminum bonding wires, which have a critical temperature of 1.175~K. (c) Inset: The total decay rate from 0.2~K to 1~K, the measurement range of Fig.~\ref{fig:temp_dep_gamma_p}.}
    \label{fig:Fig_LC_temp}
\end{figure}

Although the LC resonator has an overall temperature dependence due to the the temperature dependent characteristics of the various surface mount components comprising it,  these changes are negligible compared with the temperature dependence of $\gamma_\mathrm{p}$ and $\Delta$ observed between 0.17~K and 0.9~K in Fig.~~\ref{fig:temp_dep_gamma_p}.  Fig.~\ref{fig:Fig_LC_temp} shows fitting parameters of the bare LC resonator extracted from frequency sweeps at different temperatures. Over the temperature range from 0.17~K to 0.9~K, the resonance frequency changes by less than 0.07~MHz, and the total decay rate $\kappa/2\pi$ changes by only 0.02~MHz. 

\input{plasmon.bbl}

\end{document}

%% file: plasmon.bbl
%

%% file: main.bbl
\begin{thebibliography}{50}
\bibitem [{not()}]{note:minus_sign}%
  \BibitemOpen
  \href@noop {} {}\bibinfo {note} {Although phasor analysis typically assumes a time dependence of \( e^{i\omega t} \), here we follow the convention used in Eq.~\ref{eq:conductivity_cont} and adopt \( e^{-i\omega t} \) as the time dependence. This choice is made to ensure consistency with the input--output theory employed in the analysis. As a consequence, the imaginary unit \( i \) appears with an overall sign change throughout the expressions.}\BibitemShut {Stop}%
\end{thebibliography}
